\documentclass[5p,times]{elsarticle}
\pdfoutput=1
   
\usepackage{graphicx,miller,bm,booktabs,float,nicefrac,url,breakurl,color,microtype}
\usepackage{siunitx}[=v2]
\usepackage[version=4]{mhchem}
\usepackage{hyperref}
\usepackage[export]{adjustbox}
\usepackage{upgreek}

\hypersetup{breaklinks,hidelinks}

\DeclareMathAccent{\wtilde}{\mathord}{largesymbols}{"65}

\journal{Acta Materialia}

\begin{document}

\begin{frontmatter}

\title{A micromechanical study of  heat treatment induced hardening  in $\alpha$-brass}

\author[a,b]{Jonathan Birch}
\author[b,c]{Emily Jenkins}
\author[a]{Anastasia Vrettou}
\author[a]{Mohammed Said}
\author[c]{Himanshu Vashishtha}
\author[d]{Thomas Connolley}
\author[a]{Jeff Brooks}
\author[c]{David M. Collins}
\ead{dmc51@cam.ac.uk}
\affiliation[a]{organization={School of Metallurgy and Materials, University of Birmingham},
        addressline={Edgbaston}, 
        city={Birmingham},
        postcode={B15 2TT}, 
        country={United Kingdom}}
\affiliation[b]{organization={BAE Systems},
        addressline={Radway Green}, 
        city={Crewe},
        postcode={CW2 5PJ}, 
        country={United Kingdom}}
\affiliation[c]{organization={Department of Materials Science and Metallurgy, University of Cambridge},
        addressline={27 Charles Babbage Rd}, 
        city={Cambridge},
        postcode={CB3 0FS}, 
        country={United Kingdom}}
\affiliation[d]{organization={Diamond Light Source Ltd.},
        addressline={Harwell Science and Innovation Campus}, 
        city={Didcot},
        postcode={OX11 0DE}, 
        country={United Kingdom}}
        
\begin{abstract}
The mechanisms that govern a previously unexplained hardening effect of a single phase Cu-30wt\%Zn $\alpha$-brass after heating have been investigated. After cold-work, the alloy possesses an increased yield strength and hardening rate only when heat treated to temperatures close to 220$^\circ$C, and is otherwise softer. Crystallographic texture and microstructure, explored using electron backscatter diffraction (EBSD), describe the deformation heterogeneity including twin development, as a function of heat treatment. When heated, an increased area fraction of deformation twins is observed, with dimensions reaching a critical size that maximises the resistance to dislocation slip in the parent grains. The effect is shown to dominate over other alloy characteristics including short range order, giving serrated yielding during tensile testing which is mostly eliminated after heating. In-situ X-ray diffraction during tensile testing corroborates these findings; dislocation-related line broadening and lattice strain development between as worked and heated $\alpha$-brass is directly related to the interaction between the dislocations and the population of deformation twins. The experiments unambiguously disprove that other thermally-induced microstructure features contribute to thermal hardening. Specifically, the presence of recrystallised grains or second phases do not play a role. As these heat treatments match annealing conditions subjected to $\alpha$-brass during deformation-related manufacturing processes, the results here are considered critical to understand, predict and exploit, where appropriate, any beneficial process-induced structural behaviour.

\end{abstract}


\end{frontmatter}

\section{Introduction}

{\color{black}

Applications requiring the use of $\alpha$-brass in intricately shaped parts often require the use of mechanical pressing or spinning operations. In order to fabricate such shapes in an efficient manner, the material must possess a balance between favourable ductility, work hardening rates and high tensile strength. For shapes requiring significant levels of strain, it is often necessary to anneal the material part way through the manufacturing cycle.

When annealed, $\alpha$ phase Cu-Zn alloys are heated to temperatures where no phase changes are expected; the binary Cu-Zn phase diagram reports no transformations within the $\alpha$ phase field up to the solidus \cite{ASM}. This has been historically debated due to the presence of an anomalous hardening phenomena, found to exist when $\alpha$-brass under certain material pedigrees such as the Cu/Zn ratio, microstructure and deformation state, is heated to temperatures within the range $\sim$250$^\circ$C \cite{Lee1983} to $\sim$300$^\circ$C \cite{Mishima1948,Hasiguti1950}. To rationalise such observations, including a broader discussion of $\alpha$-brass deformation phenomena \cite{Clareborough&Loretto1960}, a speculated low temperature phase field was included within the $\alpha$ phase of the Cu-Zn binary system \cite{Elliot1965}. However, no direct evidence has, to date, corroborated this. The work of Shinoda \cite{Shinoda1960} provided a rigorous description of the $\alpha$+ $\beta'$ phase boundary curvature,  disproving the likelihood of transformations within the $\alpha$ phase field. 

Various studies have since explored the potential causes of anneal hardening in the Cu-Zn alloy system \cite{Lee1983, Clareborough&Loretto1960,Fang2020}. Changes to the physical properties from annealing may give rise to chemical short range order (SRO), physical SRO, or diffusion of structural defects \cite{Egami1978}.  SRO is defined here as the diffusion of elements to energetically preferred lattice sites, a precursor for long-range ordering of a thermodynamically favoured phase. Candidates include the formation of Cu$_3$Zn \cite{Hong2014} or Cu$_5$Zn$_8$ \cite{Gourdon2007} $\gamma$ phase superstructure, however, the latter is an unexpected formation within the $\alpha$ phase  \cite{Obenhuber1987}. Instead, its formation is possible via the transformation $\beta' \rightarrow \alpha + \gamma$ at the eutectic composition on the  Cu-Zn phase diagram at $\sim$250$^\circ$C \cite{Shinoda1960}. Considering the diffusivity of Zn in Cu is low, thermodynamic equilibrium can only be achieved after extensive time at this temperature. Such phase transformations are known to be accelerated if the initial material contains cold work \cite{Shinoda1932}. In the absence of direct observations of a secondary phase, SRO in $\alpha$-brass has been proposed to rationalise anneal hardening \cite{Obenhuber1987}. The activation temperature to initiate SRO, and hence the energy, is lower with increased zinc composition \cite{Youssef1972}. This alters the internal friction of $\alpha$ brass, reaching a maximum for Cu-30wt\%Zn at  $\sim$245$^\circ$C. SRO is also promoted with increased vacancy density, which is proportional to the diffusion of Zn; a greater solute concentration provides more opportunities for ordering. Physical SRO occurs when the relative positions of the atoms change \cite{Egami1978}. This incorporates structural relaxation when heated due to the redistribution or transformation of defects. Structural defects, in context to $\alpha$-brass, comprise substitutional or interstitial alloying elements, deformation or annealing twins. 

The deformation of $\alpha$-brass has several interesting characteristics including repeated yielding, abnormal work hardening and strain rate sensitivity \cite{Izumi1970}. During tensile loading, the deformation mechanisms are strongly dependant on crystal orientation and the propagation of L\"uder bands, governed by the magnitude of the stacking fault energy \cite{Neuhauser1975,Kocks2003,Zehetbauer1993}.  Primarily for CuZn $\alpha$-brass, the dislocation motion is octahedral slip, fine twinning or by the movement of shear bands. Beyond the role of the primary alloy constituents, Cu and Zn, trace elements may promote structures within the Cu-Zn lattice, or change the thermodynamic stability of the primary phases via lattice parameter modification. Where present, resultant localised residual stress impede dislocation slip, leading to hardening. Many elements are capable of reducing the ductility and increasing the hardness of $\alpha$-brass; notably arsenic, antimony, iron and lead \cite{Jevons1940, Davies2005}. 

Plastic instability, here in the context of dislocation motion, inhibited by pinning effects are dependent on crystal orientations as well as interactions with twinned structure \cite{Lee1983}. A ductility reduction has been associated with the density of stacking faults that act to pin the dislocations, resulting in sessile slip bands, preventing further deformation and promoting early fracture \cite{Victoria2000}. During tensile loading, $\alpha$-brass is also known to possess serrated yielding. As described by Rodriguez \cite{Rodriguez1984} and Chen \cite{Chen2018}, the amplitude of the serrations are sensitive to temperatures and strain rate. The behaviour is attributed to dislocations becoming pinned/unpinned by solute atoms. Texture also plays a role in the mechanics of $\alpha$-brass. During low levels of strain, twins form that modify the texture, comprising both copper $\{112\}\langle11\bar{1}\rangle$ and brass $\{110\}\langle1\bar{1}2\rangle$ fibres  \cite{Eldanaf2000}. Cross slip is said to be the fundamental process behind the texture transitions in FCC (face centered cubic) structures, through its link with the deformation twinning process \cite{Leffers2008}. 

At the microstructural length scale, the formation of annealing twins and deformation twins are well reported to inhibit the dislocation motion via the creation of twin boundaries \cite{Beyerlein2014}. In A1 structured CuZn $\alpha$-brass, twin formation relies on the passage of $\{111\}\langle11\bar{2}\rangle$ partial dislocations with the formation of a stacking fault. Twin generation is favoured from the low stacking fault energy of brass \cite{ZHANG2012871}. When present at the nano-scale, they dramatically increase strength and hardness \cite{Zhang2006,Beyerlein2014}. The higher mechanical strength is linked to the interactions of the dislocations with coherent twin boundaries; the nano-scale distance between the twin boundaries limits slip distance within an individual lamellae. Extensive dislocation pile up results, thus hardens the alloy.

A non-linear relationship exists between the thickness of a twin lamellae and the material strength \cite{Beyerlein2014}. A peak flow stress exists for a given lamellae thickness with respect to the grain size \cite{Lu2009, Li2010, Zhang2012}, with any macroscopic mechanical softening due to the transition in the deformation mechanisms at a critical lamellea thickness \cite{Wu2011}. Above the critical thickness, cross slip leads to Lomer-Cottrell locks, restricting further slip.  With lamellae below a critical size, cross slip does not occur. Instead, dislocation steps form on the twin boundaries; these act as nucleation sites for further dislocation slip, creating an additional pathway to mediate plasticity. Material softening may also occur from detwinning. Here, twin constriction may occur from surrounding stresses, leading to twin removal which has an associated favourable energy reduction. \cite{Wang2010}. Alternatively, the increase in density of preexisting dislocations at the twin boundaries as the twin area drops, create source of mobile dislocations for further slip during deformation. 

Prior work has shown that numerous mechanical and microstructure features affect the macroscopic properties of brass, however, there is a lack of clarity of their sensitivity to the deformation response after heating, particular with prior cold work. Considering these aspects, this study aims to identify the $\alpha$ brass characteristics that give rise to the anomalous strengthening effect.  Here, measurements are made at the meso- and microscale to understand the relationship between microstructure, crystallography and deformation mechanics.

\begin{figure}
\centering
    \includegraphics[width=90mm]{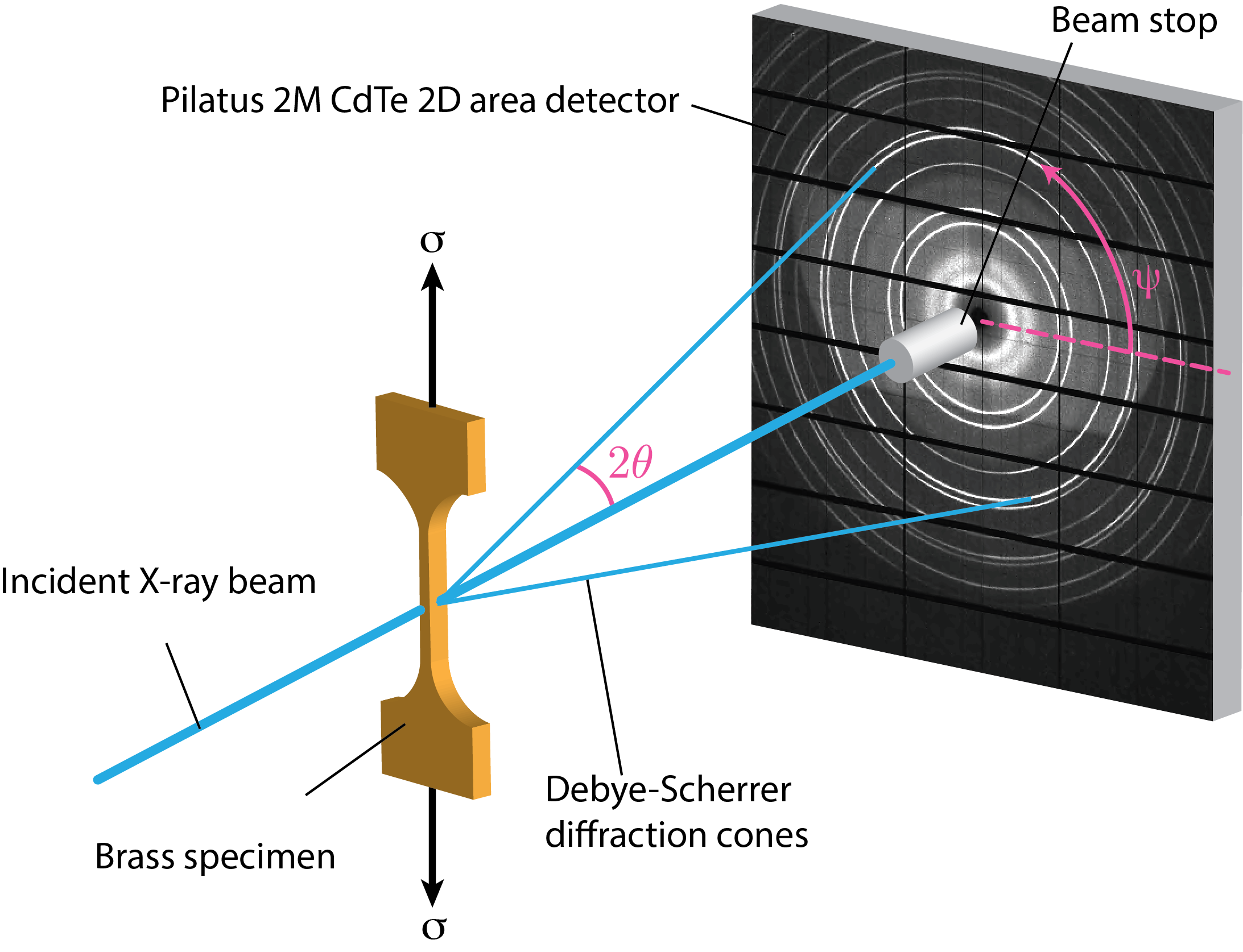}
    \caption{Configuration of the diffraction experiments during in situ deformation of brass specimens on the I12 beamline, Diamond Light Source.}
    \label{fig:expt_diagram}
\end{figure}

\section{Experimental Method}
\subsection{Material}
The anneal hardening behaviour was examined on the $\alpha$-brass alloy, Cu-30wt\%Zn. The tested brass was received in a plate form, with the cast ingot having been rolled then  annealed using proprietary conditions. The alloy conformed to a certified composition tolerance of $\pm 1\%$ for Cu and Zn. In this study, the material was subjected to cold rolling parallel to the prior rolling direction (RD), achieving an engineering strain, $\varepsilon_{\rm RD}$ of 0.36. Experiments were performed on samples in the as-rolled condition, and following heat treatment. Samples were subjected to isothermal heat treatments for 45 minutes (hardness testing and electron microscopy characterisation), or 10 minutes (X-ray diffraction measurements} at temperatures ranging from 223$^\circ$C to 385$^\circ$C. All samples were quenched in water. These heat treatments were selected as they explore the typical process conditions of $\alpha$-brass during component fabrication.  All samples for subsequent testing were made via wire type electrical discharge machining (EDM).

\subsection{Differential Scanning Calorimetry (DSC)}
To inform the selection of heat treatments after cold-rolling, DSC testing was performed. This was conducted using a Netzsch Thermal Analysis DSC 404 C cell and 85\,$\upmu$L alumina crucibles. The mass of the test specimens sectioned were 0.134\,g. All crucibles were heated to a temperature of 600$^\circ$C for 30 minutes prior to the start of the test and allowed to cool in air. The DSC test was run from 30$^\circ$C through to 500$^\circ$C at a rate of 10$^\circ$C/minute. This test was then repeated under the same settings. To account for any thermal mass errors from the test, the sample was run against an empty crucible.

\subsection{Hardness}

Hardness testing was conducted using a Vickers hardness testing machine fitted with a diamond indenter and a 1\,kg load. The cold rolled surface of the material was polished using Brasso and a cloth prior to testing. Measurements were made on the as-rolled condition, and after isothermal heat treatments. For each sample, 5 indentations were made with a spacing of $>$4\,mm.

\subsection{Electron Backscatter Diffraction (EBSD) data collection}

Sample surfaces were ground with SiC abrasive paper before polishing down to \SI{9}{\micro\meter} using a diamond media and a mol cloth. Samples were then electro-polished using a Struers LectroPol 5 with an ASTM E1558 Group 111 electrolyte 10. The polish was performed at 5\,V for 45 seconds. EBSD was conducted immediately following preparation to avoid any surface oxidation or contamination.
EBSD examination was completed using a Zeiss Evo 10 scanning electron microscope, operated at 20\,kV, equipped with an Oxford Instruments C-Nano detector. Data collection and diffraction pattern indexing was performed using the software, AZtecHKL. For each sample, EBSD maps were collected at (1) a low magnification to assess texture, using a step size of 5.4\,$\upmu$m from a $\sim2100\times2800$\,$\upmu$m$^2$ region, then (2) a higher magnification to asses deformation structures, here using a step size of 0.76\,$\upmu$m from a $\sim590\times780$\,$\upmu$m$^2$ region.

Offline data analysis was performed on EBSD datasets, using the Matlab package, MTEX 5.6.0 \cite{Mtex}. The software was used to produce inverse pole figure maps and calculate high angle grain boundaries ($>10^\circ$ misorientation). Following the calculation of the lattice curvature tensor, an estimate of the total geometrically necessary dislocation (GND) density was also obtained, following Pantleon \cite{PANTLEON2008994}. The calculation accounts for dislocation activity on all FCC slip systems, where the Burgers vector, $b = \frac{a}{2}\langle110\rangle$. Orientation data from the EBSD datasets were also used to quantify the area fraction of deformation and annealing twins. Twin boundaries were firstly identified from a misorientation of $60^{\circ}\pm3^{\circ}$ between the parent grain and the twin itself. The area fraction is calculated from the enclosed region within the twin boundary compared to the area of parent grains. From a bimodal size distribution of the twin population, the deformation twins were determined as the subset with an area  $<100\,\upmu$m$^2$ and the annealing twins were those with $\geq100\,\upmu$m$^2$.

\subsection{Synchrotron X-ray diffraction}
A high energy diffraction experiment was performed on the I12 beamline, Diamond Light Source \cite{Drakopoulos2015828}; this experimental technique has been extensively used to explore the structural behaviour of A1 structured materials e.g. \cite{Budrovic2004, Brandstetter2006, Petegem2013, Miyajima2016}.Referring to the experimental setup shown in Fig. \ref{fig:expt_diagram}, samples were subjected to uniaxial deformation whilst collecting diffraction data, in-situ. Debye Scherrer diffraction data were acquired in a transmission geometry on a Pilatus 2M CdTe area detector ($1475 \times 1679$ pixels and a pixel size of $172\times172$\,$\upmu$m$^2$) at a frame rate of 1\,Hz ($\sim1$\,s exposure time). Calibrated using a CeO$_2$ standard, as described in \cite{Hart20131249}, a  monochromatic incident X-ray at beam energy of 79.63\,keV was used and a sample to detector distance of $\sim789$\,mm.

\begin{figure}[h!]
\centering
    \includegraphics[width=90mm]{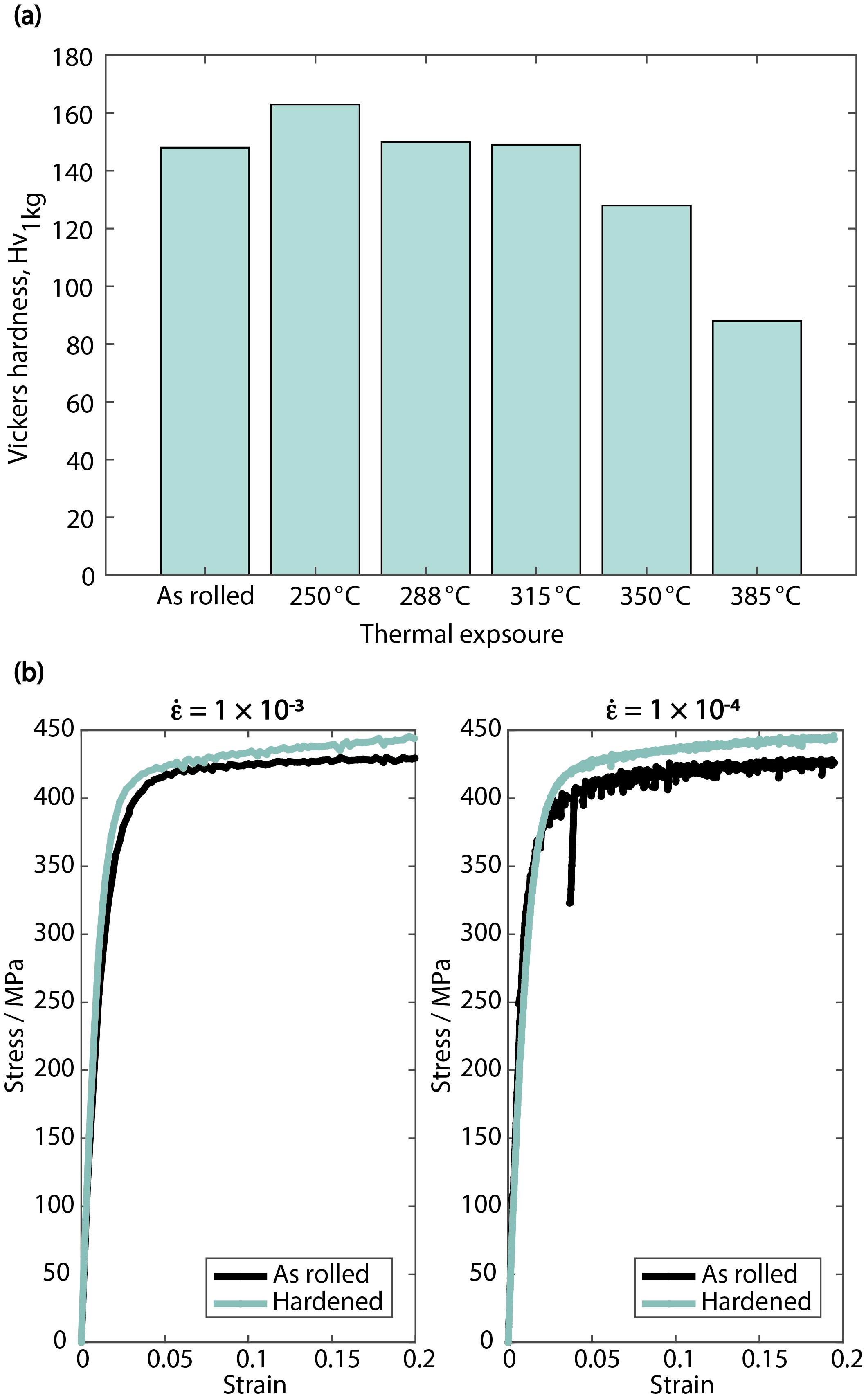}
    \caption{(a) Vickers hardness measurements for samples subjected to different annealing temperatures, and (b) Stress Strain Curves taken from the ETMT during the SXRD experiment.}
    \label{fig:MacroMechTest}
\end{figure}

Synchrotron samples were deformed to 30\% cold work via plain strain rolling, prior to being sectioned for testing with an EDM wire machine. Sample dimensions were 2\,mm in width by 1\,mm in thickness with a gauge length of 16\,mm. 
In-situ tensile testing, parallel to the $\alpha$-brass rolling direction, was completed using an Instron Electronic Thermal Mechanical Testing machine (ETMT) \cite{Roebuck2022}. Tensile loading was completed via the relative displacement of the ETMT grips; the strain rates were either $\dot{\varepsilon}=1\times10^{-3}$\,s$^{-1}$ or $\dot{\varepsilon}=1\times10^{-4}$\,s$^{-1}$.

Diffraction patterns were radially integrated into 36 equally spaced $10^\circ$ bins within the azmimuthal range $0^\circ \leq \psi < 360^\circ$. Reflections within each diffraction sector were fitted using single peaks using an in-house written method in MATLAB (i.e. \cite{VRETTOU2022143091}). Pseudo-Voigt functions were fitted to the \{111\}, \{200\}, {220} \& \{311\} reflections, recording the peak position, line profile width, integrated intensity and $\chi^2$ fitting error.

\section{Results}

\subsection{Macromechanical Behaviour}

Vickers hardness testing of $\alpha$-brass sample heat treated at different temperatures were first examined. Results are shown in Figure \ref{fig:MacroMechTest}a. From a baseline 145\,Hv in the as-rolled condition, the material hardness increases by $\sim10$\% (to 160\,Hv) after heating to 250$^\circ$C. Heating to higher temperatures sees the hardness first drop to the baseline value, becoming softer when heated to $\geq 350^\circ$C.

\begin{figure}[h!]
\centering
    \includegraphics[width=90mm]{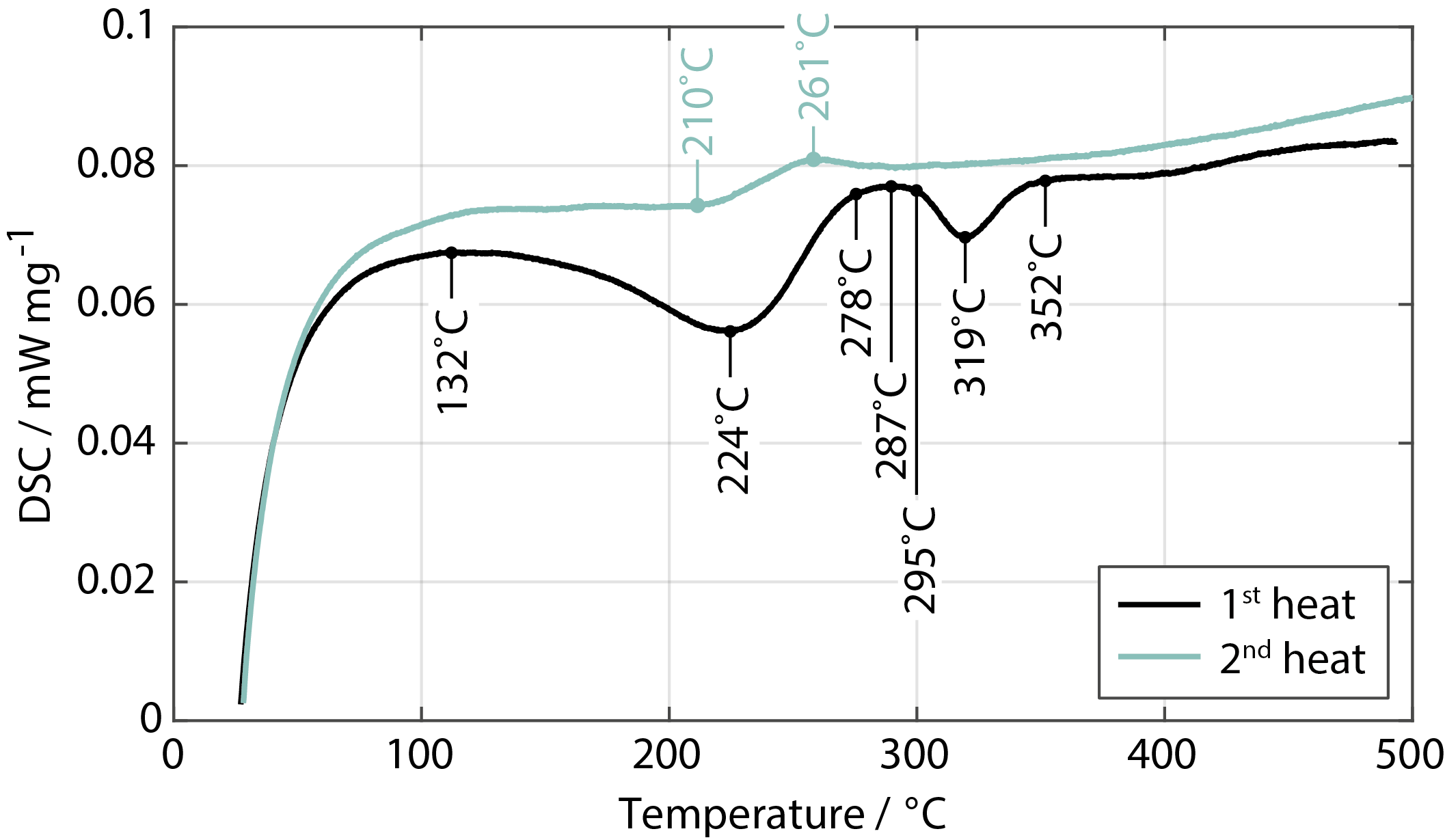}
    \caption{DSC measurements of work hardened brass specimens during a 1$^{\rm st}$ and 2$^{\rm nd}$ heating cycle.}
    \label{fig:DSC}
\end{figure}

Uniaxial tensile testing results, conducted on the as-rolled and after heat treating at (250$^\circ$C), hereon called the {\it hardened condition} are shown in Figure \ref{fig:MacroMechTest}b. At $\dot{\varepsilon}=1\times10^{-3}$\,s$^{-1}$ the hardened condition possesses (i) a small yield stress increase, and (ii) a higher hardening rate, over the as-rolled condition. At the slower strain rate, $\dot{\varepsilon}=1\times10^{-4}$\,s$^{-1}$, (iii) an increased yield stress is also evident, and (iii) serration in the plastic region is present only in the as-rolled state. 

\subsection{Differential Scanning Calorimetry (DSC)}
Given the temperature dependent behaviour evident in mechanical tests, the heat flow behaviour was assessed via DSC thermal analysis. The DSC measurements, shown in the Figure \ref{fig:DSC}, have two distinct regions of exothermic activity during sample heating (1$^{\rm st}$). A notable change is noted between 224$^\circ$C and $\sim$278$^\circ$C; this corresponds to the region that hardens the $\alpha$ brass alloy. Following the hardness data, any hardening is removed over this temperature. Heating further, a change is evident between 300$^\circ$C and 350$^\circ$C; this incurs a smaller exothermic change than the lower temperature transformation. It is well established that a measured enthalpy change can be associated with strain release processes \cite{Lu2001}, where their onset temperature is heavily dependent on the plasticity levels. This is supported by evidence here when heated for a second time; no significant exothermic features are present. Starting with a high level of cold work, recovery and recrystallisation processes must be operative, however, evidence for each is needed. 

\subsection{Microstructure Characterisation}

\begin{figure*}[t!]
\centering
    \includegraphics[width=160mm,center]{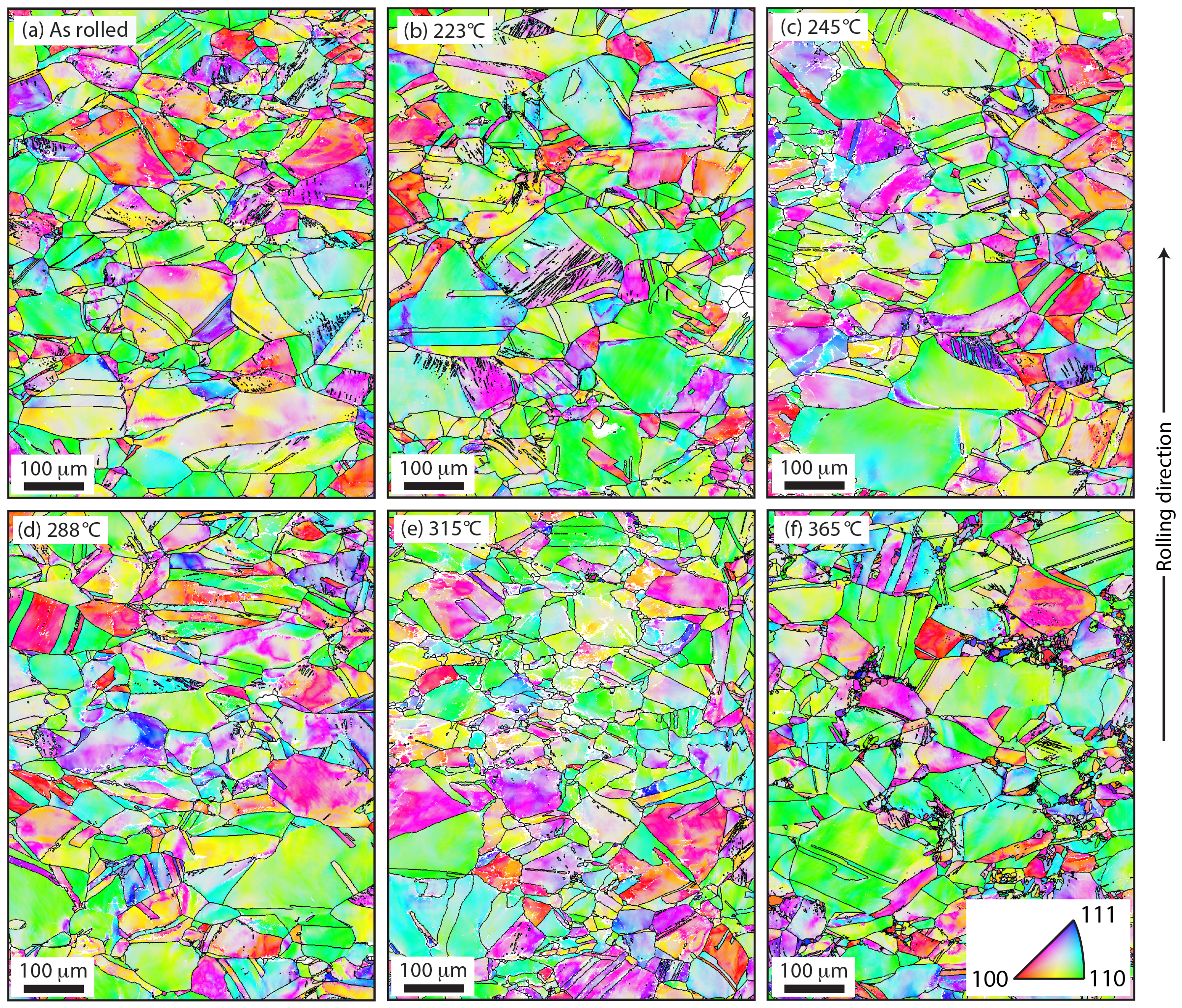}
    \caption{EBSD inverse pole figure maps with respect to the normal direction to the sheet. A map is shown for each examined condition.}
    \label{fig:EBSD}
\end{figure*}

To further explore the macroscopic strength differences between cold worked and samples that have been subsequently heat treated, microstructure characterisation was performed with EBSD. Specifically, the $\alpha$-brass was assessed in the as-rolled condition, and after being heat treated at either 223$^\circ$C, 245$^\circ$C, 288$^\circ$C, 315$^\circ$C or 365$^\circ$C. The corresponding inverse pole figure maps are shown in Figure \ref{fig:EBSD}; the orientations are plotted with respect to the normal direction to the rolled sheet. The high angle grain boundaries ($15^\circ$ minimum) are also superimposed. All datapoints are of the same phase; $\alpha$, with space group $F m {\bar 3} m$. Several features are observable from these plots: (i) annealing twins are present in all examined conditions, (ii) grains possess significant misorientation within them, this is evident after cold rolling and for all heat treated conditions, (iii) deformation twins are evident in all samples, but are most prevalent in the as rolled (Figure \ref{fig:EBSD}a) and 223$^\circ$C (Figure \ref{fig:EBSD}b) samples, and (iv) partial recrystallisation occurred in the samples heated to the highest temperatures, 315$^\circ$C (Figure \ref{fig:EBSD}e) or 365$^\circ$C (Figure \ref{fig:EBSD}f); clusters of small grains, $<10$\,$\upmu$m, are present. The recrystallisation corresponds to the exothermic change observed in the DSC measurements within this temperature range. 

\begin{figure*}[t!]
\centering
    \includegraphics[width=190mm,center]{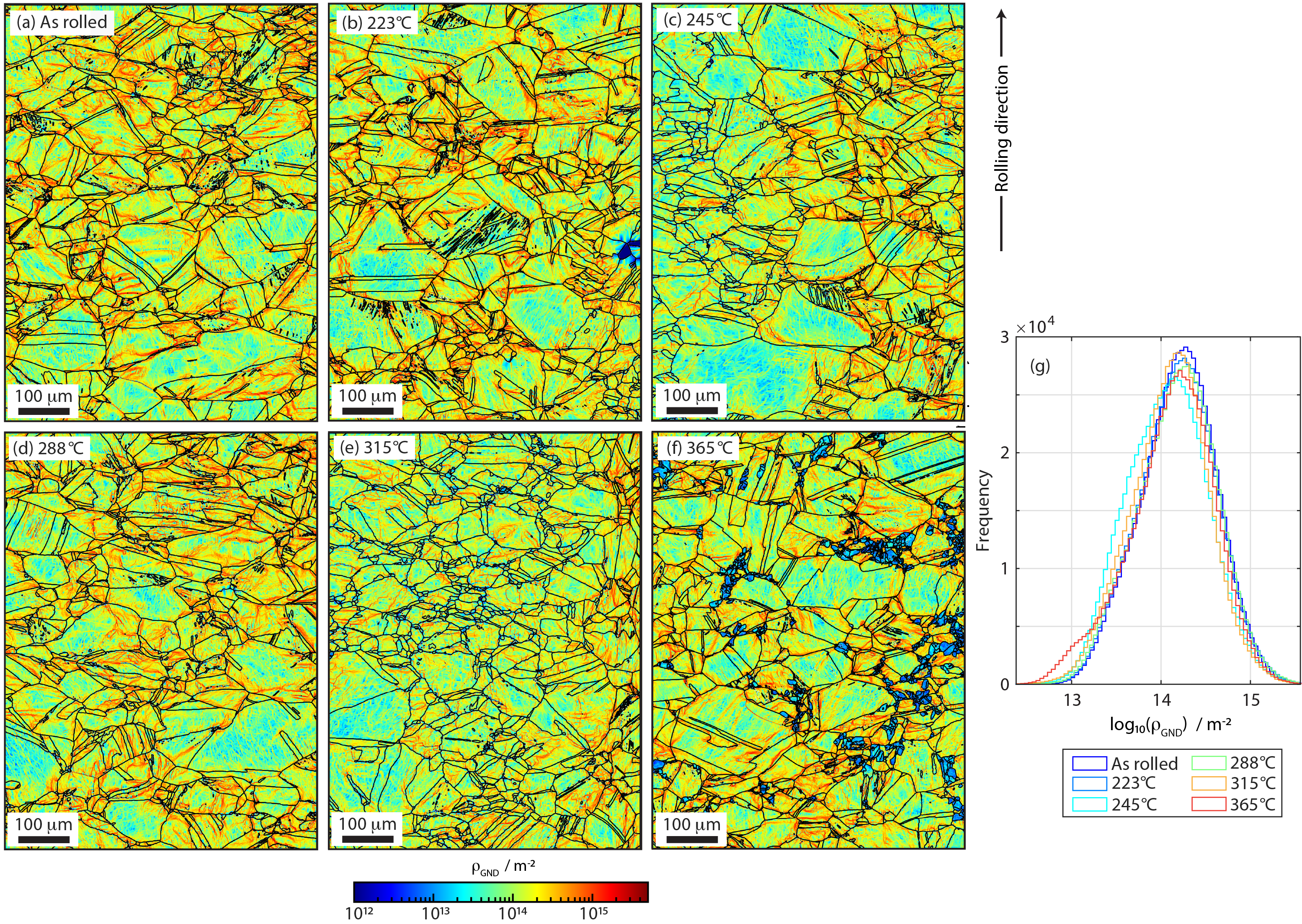}
    \caption{GND maps estimated from EBSD orientation measurements. Histograms for each of the maps are shown in (g).}
    \label{fig:GND}
\end{figure*}

Estimates of the GND densities for each of the assessed deformation conditions are shown in Figure \ref{fig:GND}. All samples have evidence of deformation structures, with higher dislocation densities closer to grain boundaries; high GND density ($>10^{15}$\,m$^-2$) bands and cellular-like structures. The largest grains having regions of lower GND density ($<10^{14}$\,m$^-2$), again characteristic for all samples. The spatial organisation of the dislocations do not highlight any standout features in the hardened states (heated to $<250^\circ$C). At elevated temperatures, $> 300^\circ$C, where an exothermic peak was observed (DSC Fig. \ref{fig:DSC}), recrystallisation is evident in the GND maps with small grains of low density ($\rho_{\rm GND}<10^{13}$\,m$^{-2}$). Plotting histograms for each map, Fig. \ref{fig:GND}g, shows a systematic decrease in $\rho_{\rm GND}$ as recovery ensues. A low density peak associated with the newly formed recrystallised grains is clear for the 365$^\circ$C sample.

\subsection{Twin Fraction}

\begin{figure*}[t!]
\centering
    \includegraphics[width=190mm,center]{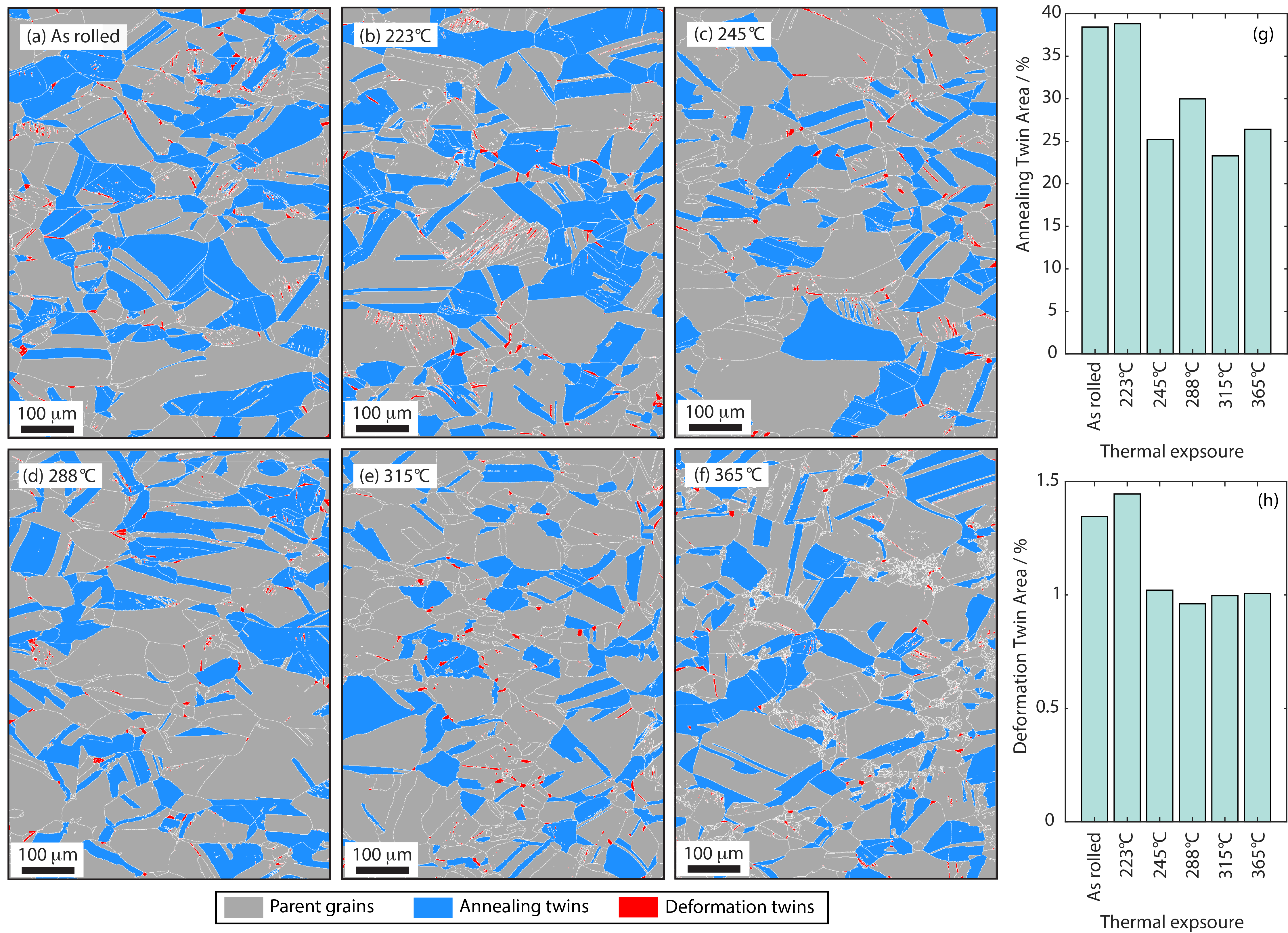}
    \caption{Grains maps (a-f) calculated from EBSD datasets for the as-rolled and heat treated conditions, coloured by parent $\$\alpha$ grain, annealing twins or deformation twin. An estimate of the corresponding annealing twin (g) and deformation twin (h) fractions are also shown.}
    \label{fig:TWINS}
\end{figure*}

Grain maps that identify the deformation and annealing twins are shown in Fig. \ref{fig:TWINS}a-\ref{fig:TWINS}f. The corresponded area fractions calculated for each twin type are reported in Fig. \ref{fig:TWINS}g-\ref{fig:TWINS}h. The annealing twin fraction notably drops for temperatures over 245$^\circ$C; evidently migration of the twin boundaries is sufficient for their elimination. Errors here arise from ambiguity within the grain as to which part is the parent or child (i.e. the twin), thus, these results are considered only as qualitatively reliable. Between the as-rolled and heating to 245$^\circ$C shows an increase in deformation twin area fraction of $\sim10$\%; this is considered as a substantial finding and corresponds to the temperature where hardening occurs. Heating to a higher temperature corresponds to a drop in deformation (and annealing) twin fraction, following the trend in hardness tests.

\subsection{Texture development}
\begin{figure}
\centering
    \includegraphics[width=72mm]{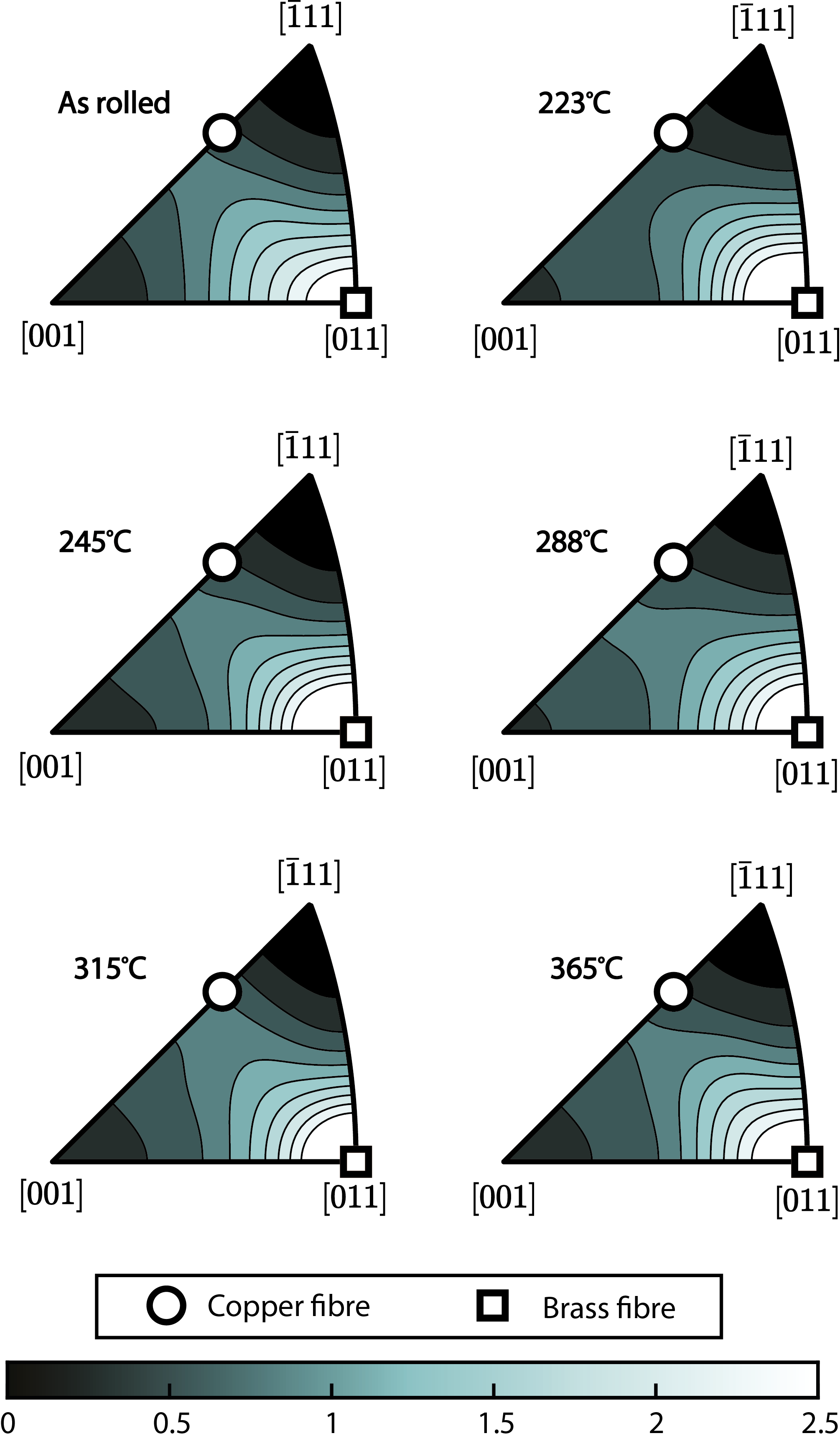}
    \caption{Inverse pole figures for for (a) As rolled material, (b) 223$^\circ$C, (c) 245$^\circ$C, (d) 288$^\circ$C, (e) 315$^\circ$C, (f) 365$^\circ$C.}
    \label{fig:IPF}
\end{figure}

An assessment of the texture from the EBSD orientation measurement is presented in Fig. \ref{fig:IPF}. All samples are shown to have a strong $\langle011\rangle$ brass fibre texture; all are approximately the same except for the sample heated to 223$^\circ$C that is weaker in the direction close to the copper fibre. This corresponds to the twin direction, $\langle112\rangle$ Correlating to prior results, this is the sample with greatest hardness and highest deformation twin fraction.

\subsection{Synchrotron X-ray Diffraction}

\begin{figure}[h!]
\centering
    \includegraphics[width=90mm]{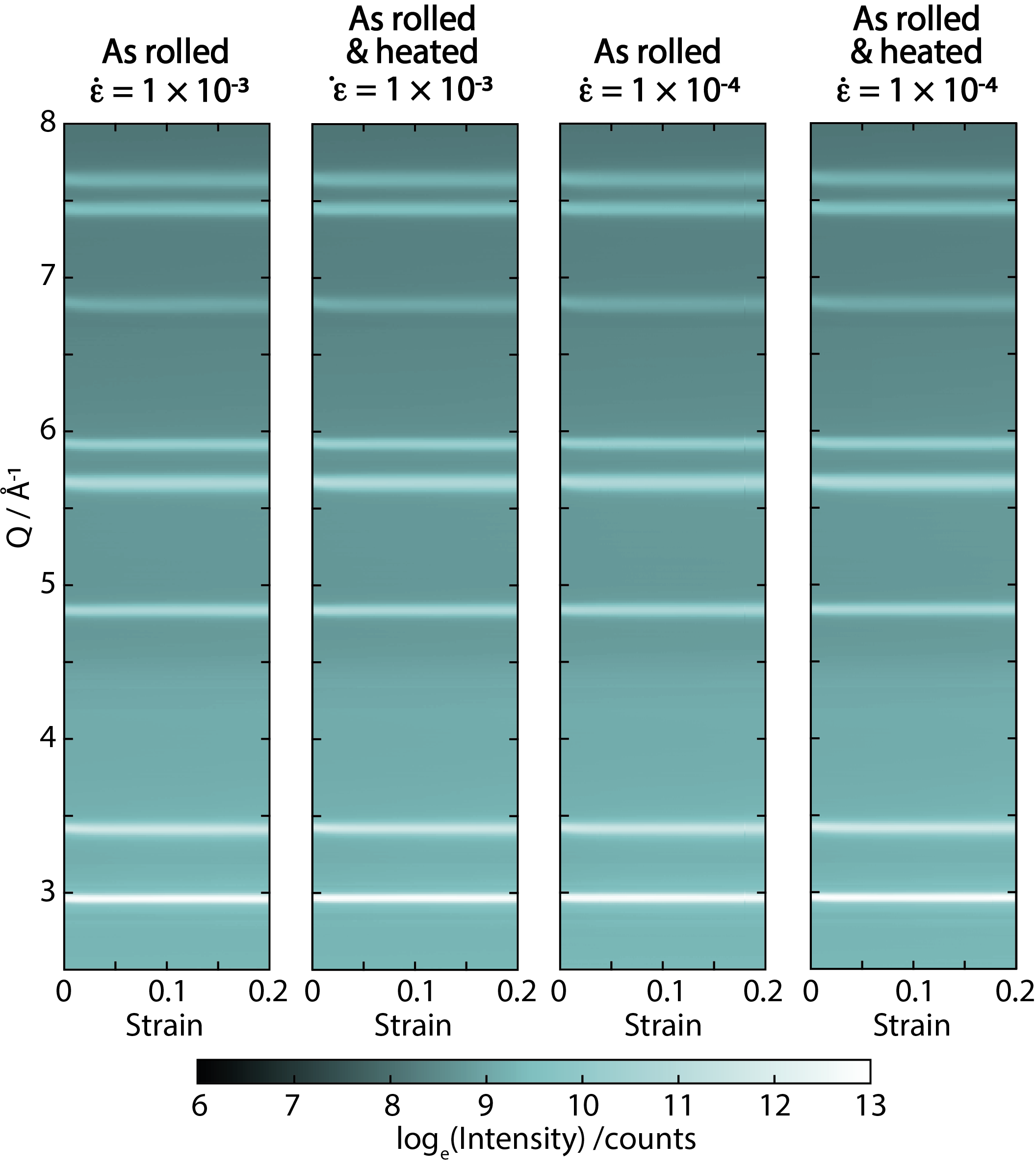}
    \caption{Diffraction data for tested conditions of CuZn30 brass when subjected to uniaxial loading. The intensity data shown are from the fully integrated Debye Scherrer rings.}
    \label{fig:RawData}
\end{figure}

\begin{figure*}[h!]
\centering
    \includegraphics[width=160mm,center]{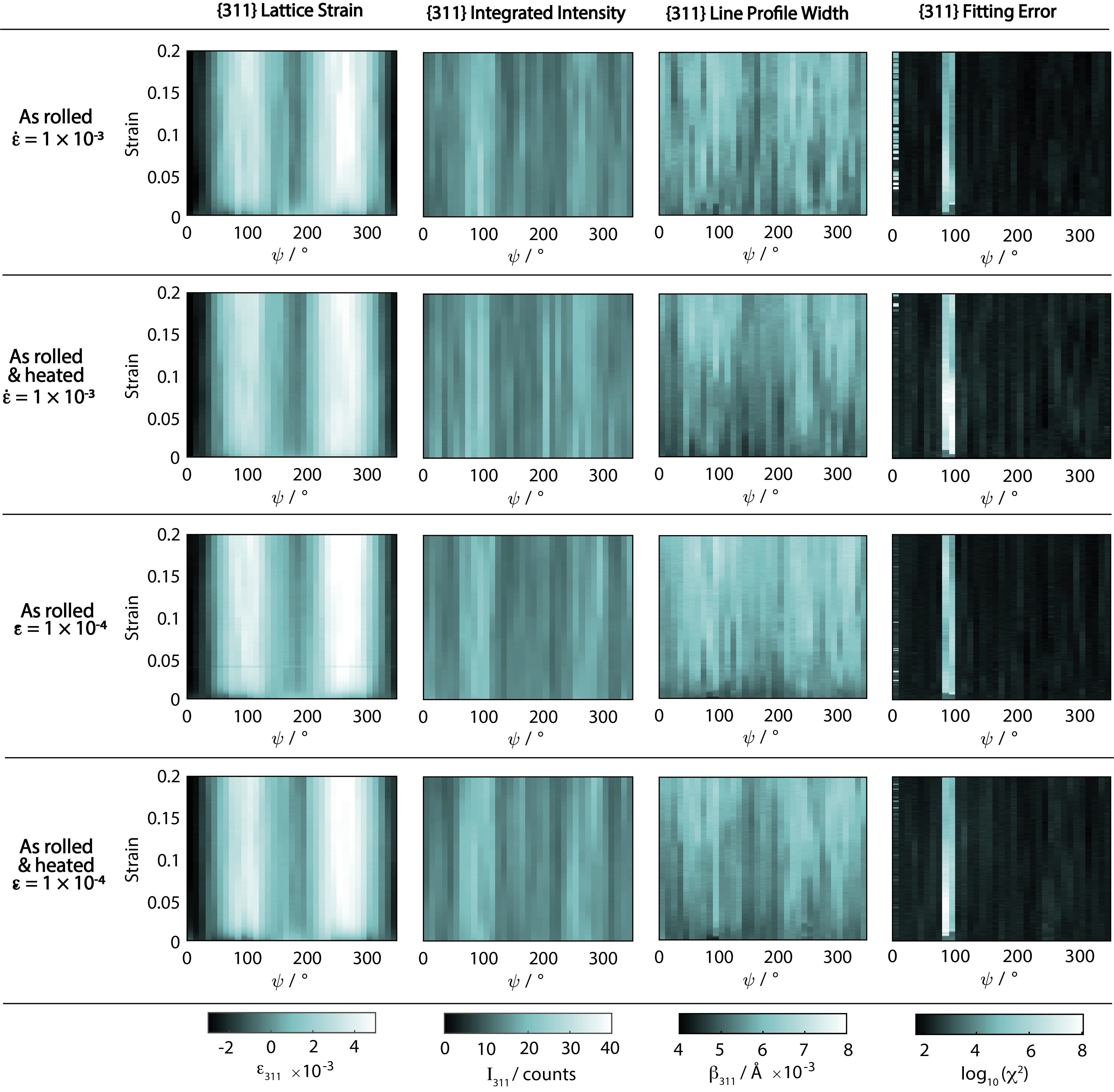}
    \caption{Fitted data for \{311\} reflection, showing the lattice strain, $\varepsilon_{hkl}$ (column 1), integrated intensity (column 2), line profile width, $\beta_{hkl}$ (column 3), and fitting error, $\chi^2$ (column 4) as a function of macroscopic strain and azimuthal orientation, $\phi$. Here, the tensile direction is $\psi=90^\circ\pm180^\circ$ and the transverse direction is $\psi=180^\circ\pm180^\circ$.}
    \label{fig:FittedData2D}
\end{figure*}

\begin{figure}[h!]
\centering
    \includegraphics[width=90mm]{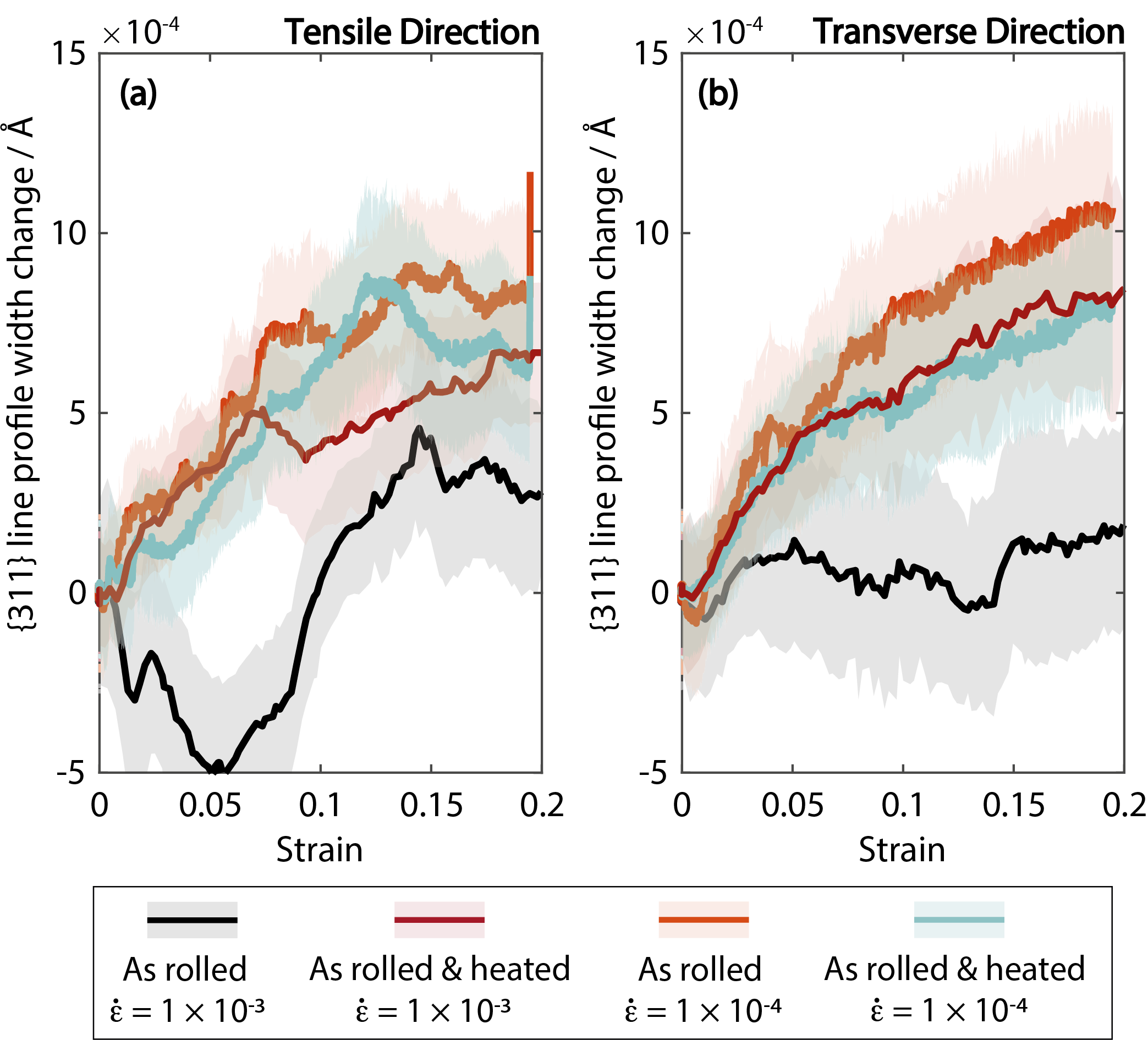}
    \caption{Diffraction data for tested conditions of CuZn30 brass when subjected to uniaxial loading. The intensity data shown are from the fully integrated Debye Scherrer rings.}
    \label{fig:WidthChange}
\end{figure}

In-situ synchrotron diffraction during tensile loading (see Fig \ref{fig:MacroMechTest}b) was performed to ascertain any crystallographic or micromechanical features that govern the hardening phenomena. The raw data acquired during testing for the as-rolled and hardened conditions, deformed at  $\dot{\varepsilon}=1\times10^{-3}$\,s$^{-1}$ and  $\dot{\varepsilon}=1\times10^{-4}$\,s$^{-1}$ are shown in Fig. \ref{fig:RawData}. This fully integrated 2D diffraction data shows evidence only of the fundamental reflections associated with the A1 structured $\alpha$-brass. There is no evidence of any additional phases nor short range order formed in any of the tested specimens. Fitting of the individual reflections, however, does reveal interesting behaviour, as shown in Fig. \ref{fig:FittedData2D} for the \{311\} peak. The lattice strain, $\varepsilon_{hkl}$ was calculated from:

\begin{equation}
\varepsilon_{hkl} = \frac{d_{hkl} - d_{hkl,0}}{d_{hkl,0}}
\end{equation}

\noindent where $d_{hkl}$ is the $d$-spacing for a $hkl$ reflection  and $d_{hkl,0}$ is reference $d$-spacing in the undeformed state (obtained when the macroscopic strain is zero). This lattice strain is plotted alongside the integrated intensity, line profile width ($\beta_{hkl}$) and fitting error ($\chi^2$) as a function of macroscopic strain and azimuthal angle, $\psi$. As plotted, there is no evidence of a strong heat treatment or strain rate dependency on lattice strain or integrated intensity response. The line profile width, however, does possess differences. In general, reflections are broader for grains families with their diffraction vector close to the tensile direction; this is intuitively expected as these grains must be subjected to the highest stresses. This is, however, true to a lesser extent for the as-rolled, $\dot{\varepsilon}=1\times10^{-4}$\,s$^{-1}$ sample where grains have similar broadening irrespective of azimuthal orientation. It is also noticeable that for samples deformed at $\dot{\varepsilon}=1\times10^{-3}$\,s$^{-1}$, the broadening rate with strain in the tensile direction is somewhat lower in the heated condition. This indirectly indicates that plasticity is accommodated by line broadening contributors, such as dislocations, in the as-rolled stated to a greater extent than the heated condition, for a given macroscopic strain. Considering this observation is not repeated for the slower strain rate tests ($\dot{\varepsilon}=1\times10^{-4}$\,s$^{-1}$), indicates the broadening contributor mechanism must be time dependent. The \{311\} broadening trends are replotted as line profiles for the tensile (Fig. \ref{fig:WidthChange}a) and transverse sectors (Fig. \ref{fig:WidthChange}b), corroborating the different behaviour possessed by the as-rolled $\dot{\varepsilon}=1\times10^{-3}$\,s$^{-1}$ condition.

\begin{figure}[h!]
\centering
    \includegraphics[width=90mm]{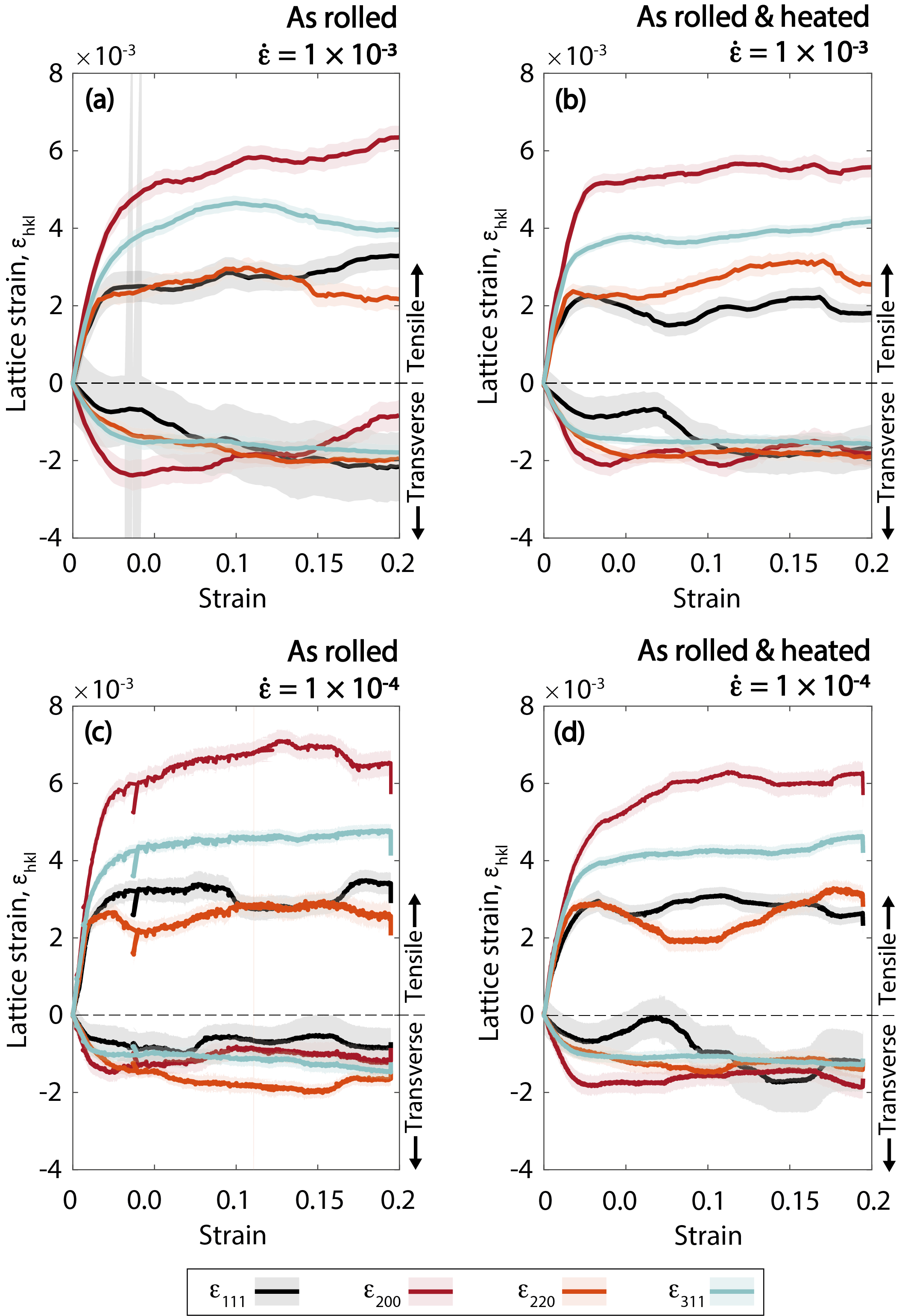}
    \caption{Lattice strain data for grain families diffracting in the (a) tensile and (b) transverse directions. The shaded region corresponds to the fitting error for each $hkl$ reflection.}
    \label{fig:LatStrain}
\end{figure}

Lattice strain data for all fitted reflections are shown in Fig. \ref{fig:LatStrain}. The ranking of the lattice strain magnitudes for each $hkl$ planes follow their respective stiffness (e.g. $\langle200\rangle$ is the elastically soft direction), as expected for a cubic material. There is limited evidence of elastic partitioning to a particular plane, however, hardening differences are present between the \{200\} and \{311\} reflections for the tests conducted at $\dot{\varepsilon}=1\times10^{-3}$\,s$^{-1}$. Whereas limited hardening is observed for these reflections in the heated state (Fig. \ref{fig:LatStrain}b), in the as-rolled condition there is an initial concomitant hardening of both lattice planes, then divergence (\{200\} hardens whilst \{311\} softens) when the strain passes 0.1. Following the EBSD results, this condition corresponds to a lower area fraction of deformation twins.

At a slower strain rate, $\dot{\varepsilon}=1\times10^{-4}$\,s$^{-1}$, serrations are observed in the lattice strain data, and to a greater extent in the as-rolled state (Fig. \ref{fig:LatStrain}c-d). The magnitude of these serrations are greater in the tensile direction, which likely govern the macromechanically observed serrations. Unlike tests at  $\dot{\varepsilon}=1\times10^{-3}$\,s$^{-1}$, the response between the as-rolled and heated conditions are similar. Together with the observed serrations, the micromechanisms that govern Cu-30wt\%Zn $\alpha$-brass are unequivocally time sensitive.

\section{Discussion}
This study aimed to understand why Cu-30\%Zn $\alpha$-brass can possess anomalous hardening when subjected to heat treatments at intermediate temperatures (circa 250$^\circ$C) following cold work. A multi-modal investigation here has amassed comprehensive and detailed datasets to describe the microstructual and structural/crystallographic characteristics at the mesoscale to microscale in an effort to isolate the features that govern the hardening. In the following narrative, candidate mechanisms are discussed in context to their likelihood of contributing to the anomalous hardening.

The role of recrystallisation is first considered. DSC results showed that there are two distinct regions of enthalpy change when cold worked $\alpha$-brass is heated to intermediate temperatures, replicating results from a prior study \cite{Clareborough&Loretto1960}. Their account of recrystallisation agrees with results in this study in Fig. \ref{fig:DSC} at 315$^\circ$C. This is directly observed from EBSD data analysis of the microstructure, (i.e. GND plots in Fig. \ref{fig:GND}) where small recrystallised grains are evident for heat treatments $\geq315^\circ$C. The  recrystallisation onset temperature is distinctly higher than any heating-induced hardening reported in a prior  \cite{Lee1983} or this study (circa $250^\circ$C). Recrystallisation is therefore not a contributor to the hardening of $\alpha$ brass. This notably contradicts a recent study by Fang \textit{et al.} \cite{Fang2020} where recrystallisation occurs at a lower temperature with associated hardening, however, this considered strain levels far higher than the present study. High strain is known to reduce the energy required for diffusion of structural defects \cite{Shinoda1932}, promoting an earlier onset of recrystallisation. 

In the vicinity of 200-250$^\circ$C, recovery of defects is expected; this corresponds to the endothermic region measured via DSC. However, GND densities showed limited change for samples heated to these temperatures, indicating that mechanisms related to dislocation recovery is not affecting the hardening response. For the postulated initiation and formation of short or long range ordering within this temperature range \cite{Clareborough&Loretto1960}, the latter is not evident. The synchrotron X-ray diffraction measurements did not show the presence of a secondary phase nor crystallographic changes that would explain anneal hardening. Short range order cannot be eliminated, with an associated process possibly contributing to the endothermic change seen here. The growth of deformation twins is, however, clear at the microstructure length scale from EBSD measurements for deformed samples heated to different temperatures. A significant contributor of the endothermic activity measured via DSC within the 200-250$^\circ$C range is therefore due to the recovery energy of twin annihilation and/or detwinning \cite{Wang2010}.

Twinning is deemed to play a key part in the strengthening of the material by heating to intermediate temperatures, where a higher measured twin fraction (annealing \& deformation) coincides with a higher hardening rate, yield stress and Vickers hardness (Fig \ref{fig:MacroMechTest}). The measured results indicated a peak deformation twin area exists after heating, however, this finding should be considered in context to the spatial resolution of the EBSD dataset (recall the step size of 0.76\,$\upmu$m). This dimension represents the narrowest twin that can be detected, hence the annealing twin fraction strictly indicates the fraction above this threshold. Any twins smaller than this will not be observed. As the twins are evidently thickening, it is likely many fine twins exist below the detection limit in the as-rolled state. Here, the deformation twin fraction is strictly defined as the observable range in context to the EBSD data collection conditions.

\begin{figure}[h!]
\centering
    \includegraphics[width=90mm]{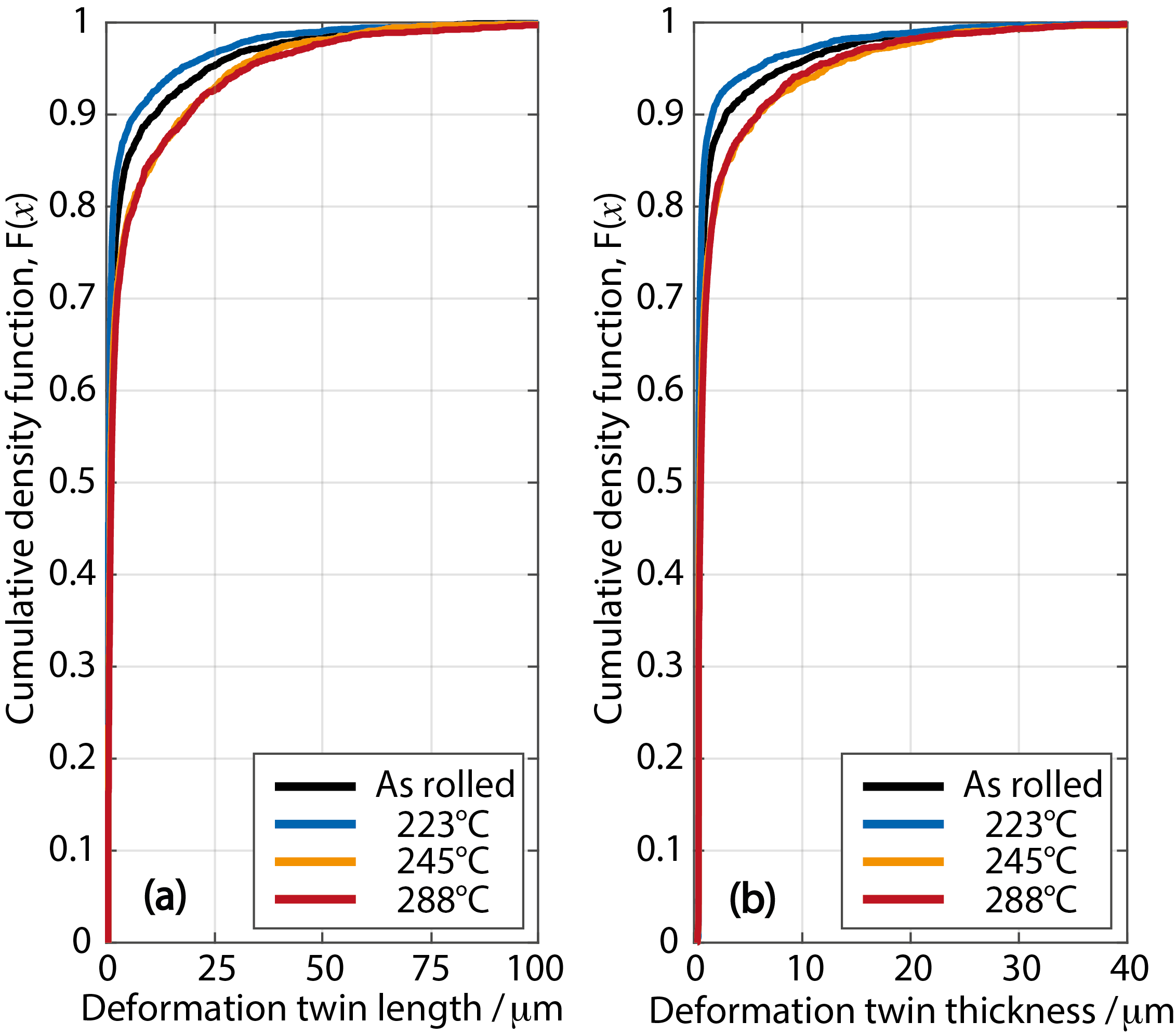}
    \caption{Estimated (a) length and (b) thickness of deformation twins for the as rolled and selected heated treatments.}
    \label{fig:TWIN_dims}
\end{figure}

A size analysis of the measured deformation twins from EBSD datasets has been performed, as shown in Fig. \ref{fig:TWIN_dims}. Using MTEX \cite{Mtex}, individual deformation twins were approximated as ellipsoids, giving a distribution of twin length Fig. \ref{fig:TWIN_dims}a and widths Fig. \ref{fig:TWIN_dims}b; these are plotted as a cumulative distribution function. Here, the conditions where recrystallisation occurs have been omitted. From the initial as-rolled condition, the population of deformation twins present are narrower and shorter in length after heating to 223$^\circ$C. Heating to higher temperatures shows the twin population is coarser; longer and thicker in all cases. Correlating these observations to the slow deformation of cold worked material, it is proposed that yield and hardening must increase when the twin lamellae are close to a critical size. Following Wu \textit{et al.} \cite{Wu2011} the twin itself acts as an obstacle, which at a critical size, will give rise to a maximum resistance to slip. Slow deformation of the cold worked material ($\dot{\varepsilon}=1\times10^{-4}$\,s$^{-1}$) again has twin lamellae below the critical size, explaining the higher yield stress, however, this does not explain the serrated yielding. Evidently the serrated yield occurs only when the deformation rate is lower, and is more pronounced prior to a heat treatment. Such serrations are consistent with SRO (i.e. \cite{Rodriguez1984,Chen2018}), which indicates that some level of SRO is present in the as-rolled state. However, as the serration amplitude reduces when heated, but the macroscopic yield strength increases, the anomalous hardening is unlikely to be governed by SRO. Indeed, the endothermic process at $\sim$224$^\circ$C shown on the DSC measurements, could be partly related to SRO being eliminated in the work hardened alloy. Instead, deformation twinning is the primary candidate to explain the effect.

The inverse pole figure plots, Fig. \ref{fig:IPF}, show the hardening alloy (223$^\circ$C) is correlated with a small texture change; here a weakening of the copper fibre is observed. This is explained by the increased deformation twin fraction, which has been previously reported to result in a texture change \cite{Leffers2008}. Given the change is small, the texture change is unlikely to govern the hardening response; no subsequent texture change was observable from X-ray diffraction measurements during tensile loading, indicating grain orientation has not affected the material response at a micromechanical length scale.

A key finding from the X-ray diffraction results was a difference in line broadening, as a function of strain, between as-rolled and heated states (see Fig. \ref{fig:WidthChange}, $\dot{\varepsilon}=1\times10^{-3}$\,s$^{-1}$). Given that broadening is caused by the presence of inhomogeneous microstrain and/or structural defects \cite{Brandstetter2006, Petegem2013,Klug1974}, the results can be used to infer dislocation behaviour. The trends imply that the dislocation density accumulates more readily in the heated condition; this corresponds with a material condition with the highest volume fraction of deformation twins, with the smallest size and with any SRO effects diminishing (inferred from serrated yielding at $\dot{\varepsilon}=1\times10^{-3}$\,s$^{-1}$). It is proposed that the small deformation twins themselves act as nucleation sources; this is corroborated by dislocation steps developing on pinned twin boundaries \cite{Wu2011}) leading to an increased dislocation density, and consequently a higher work hardening rate. At the crystallographic length scale, the resultant lattice strains are different between as rolled and heated conditions, reinforcing the claim that the slip resistance has been modified.

Evidence has been found that the leading contributor to the anomalous hardening effect in $\alpha$-brass is the development and growth of deformation twins. As the size of the twin lamellae has been noted to affect the strength of materials in a non linear fashion, it is proposed that critical twin size provides an obstacle that imposes a maximum resistance to gliding dislocations, as well as the twin boundaries providing an effective source for their nucleation. Such dislocation pinning is not as potent when the twin size and associated volume fraction within the critical strengthening effect range differ.

\section{Conclusion}

A multimodal study, including X-ray synchrotron diffraction, EBSD, DSC and mechanical testing has been performed on Cu-30wt.\%Zn $\alpha$-brass to elucidate why the material possesses an anomalous hardening when subjected to an annealing heat treatment. The work has revealed several key characteristics that contribute to this behaviour. A summary of the findings are as follows:

\begin{enumerate}

    \item{The yield strength, hardening rate and hardness of cold worked $\alpha$-brass increases after an intermediate heat treatment, around 220$^\circ$C, but diminishes when subjected to a higher temperature. The tensile behaviour is shown to be strain rate sensitive, possessing serrated yielding for slow strain rates, with serration magnitude reducing after heating.}

    \item{The hardened condition of $\alpha$-brass has a fine dispersion of deformation twins of area fraction greater than the as-worked state. The anomalous hardening can be explained by twins exhibiting a critical size that maximise the resistance to dislocation motion. This is supported by X-ray diffraction results which indicate the dislocation density increases more rapidly in the presence of the deformation twins.}

    \item{The geometrically necessary dislocation density of the as worked $\alpha$-brass was seldom affected by heat treatments, illustrating that negligible recovery had taken place and is therefore independent to the hardening phenomena. Short range order explains the serrations in the as-rolled condition, particularly  at the lower of the two strain rates studied. Reduced serrations are observed after heat treatment, so short range order cannot explain the anomalous hardening effect which occurs as a result of heating.} 

    \item{A change in texture is observed; the ratio between brass and copper texture differs when $\alpha$-brass is heated into the hardened condition. The change is considered incidental from the deformation twin development, and does not affect subsequent tensile behaviour.}
    
\end{enumerate}

\section{Acknowledgments}

The authors acknowledge Diamond Light Source for time on Beamline I12 under proposal MG29061.
BAE Systems systems are also acknowledged for their in-kind support for this study, and to the University of Manchester at Harwell for access to the ETMT.

\typeout{}

\bibliographystyle{elsarticle-num-names}

\end{document}